\documentclass[aps,prl, reprint,superscriptaddress, preprintnumbers]{revtex4-1}
\usepackage[english]{babel}
\usepackage{amsmath,amscd}
\usepackage{wasysym}
\usepackage{graphicx}
\usepackage{caption}
\usepackage{tikz}
\usepackage{amssymb}

\pdfoutput=1 

\def\be{\begin{equation}}
\def\ee{\end{equation}}
\def\bea{\begin{eqnarray}}
\def\eea{\end{eqnarray}}

\def\a{\alpha}
\def\b{\beta}
\def\g{\gamma}
\def\d{\delta}
\def\m{\mu}
\def\n{\nu}

\def\l{\lambda}

\def\s{\sigma}

\def\bma{\begin{pmatrix}}
\def\ema{\end{pmatrix}}

\def\bi{\begin{itemize}}
\def\ei{\end{itemize}}

\begin{document} 
	\preprint{IFT-UAM/CSIC-15-112}
	\preprint{FTUAM-15-34}
	\preprint{FTI-UCM/161}
	
\title{\boldmath Weyl invariance with a nontrivial mass scale.}
\author{Enrique \'Alvarez}
\email{enrique.alvarez@uam.es}
\affiliation{ Instituto de F\'{\i}sica Te\'orica, IFT-UAM/CSIC, Universidad Aut\'onoma, 28049 Madrid, Spain}
\affiliation{Departamento de F\'{\i}sica Te\'orica, Universidad Aut\'onoma de Madrid, 28049 Madrid, Spain}
\author{Sergio Gonz\'alez-Mart\'{\i}n}
\email{sergio.gonzalez.martin@csic.es}
\affiliation{ Instituto de F\'{\i}sica Te\'orica, IFT-UAM/CSIC, Universidad Aut\'onoma, 28049 Madrid, Spain}
\affiliation{Departamento de F\'{\i}sica Te\'orica, Universidad Aut\'onoma de Madrid, 28049 Madrid, Spain}

\begin{abstract}
A theory with a mass scale and yet Weyl invariant is presented. The theory is not invariant under all diffeomorphisms but only under transverse ones. This is the reason why Weyl invariance does not imply scale invariance in a free falling frame. Physical implications of this framework are discussed.
\end{abstract}








\maketitle
\flushbottom

\section{Introduction}
It is often stated that conformal invariance prevents  a mass scale to appear in any quantum field  theory \cite{Iorio}. When gravitation is dynamical, conformal invariance means Weyl invariance, that is, invariance of the theory under Weyl rescalings
\be
g_{\m\n}(x)\quad\rightarrow \quad\Omega^2(x) g_{\m\n}(x)
\ee
This cherished belief is not true anymore when the theory is not invariant under the full group of diffeomorphisms, $\text{Diff}$ but only under the volume preserving subgroup, which corresponds to transverse generators, $\text{TDiff}$. The action of unimodular gravity \cite{AlvarezGMHVM}, for example, reads

\bea\label{UG}
&&S_{UG}\equiv \int d^n x~{\cal L}_{UG}\equiv \nonumber\\
&&-M_P^{n-2}\int |g|^{1/n}\left(R+{(n-1)(n-2)\over 4 n^2}{g^{\m\n}\nabla_\m g\nabla_\n g\over g^2}\right)
\eea
and involves an explicit mass scale.
This action is not $\text{Diff}$ invariant as advertised; its symmetry is here reduced to
\be
H\equiv \text{TDiff}\ltimes \text{Weyl}
\ee

 The reason as to how this is at all possible is that when the theory is invariant under $\text{TDiff}\subset\text{Diff}$, even though the theory is Weyl invariant, the flat space limit is not necessarily scale invariant, because the Weyl dimension of the measure does not coincide with the scale dimension of $d^n x$.
 

Consider, for example, the $\text{TDiff}\ltimes \text{Weyl}$ coupling of a scalar field with gravity. There are two options. The first one is to consider matter as a Weyl singlet. The action reads 
\be\label{TDS}
S=\int d^n x~\bigg\{|g|^{1/n}g^{\m\n}\nabla_\m\phi\nabla_\n\phi-V(\phi)\bigg\}
\ee
The other option is to assign the usual  nontrivial conformal weight to the scalar, that is
\be
\phi\longrightarrow \Omega^{-1}~\phi
\ee
In that case, we only have to realize that the object
\be
|g|^{1\over 2n}~\phi
\ee
is a Weyl singlet, so that
\bea\label{TDSS}
&&S=\int d^n x~\bigg\{|g|^{1/n}g^{\m\n}\nabla_\m\left(g^{1\over 2n}\phi\right)\nabla_\n\left(g^{1\over 2n}\phi\right)-V(g^{1\over 2n}\phi)\bigg\}=\nonumber\\
&&\int d^n x~\bigg\{|g|^{2/n}~\left({1\over 4 n^2}\left({\nabla g\over g}\right)^2+{1\over n}{\nabla g.\nabla\phi\over g}+\left(\nabla\phi\right)^2\right)-V(g^{1\over 2n}\phi)\bigg\}\nonumber
\eea
is again Weyl invariant. In particular the mass term
\be
-{m^2\over 2} |g|^{1\over n}\phi^2
\ee
is perfectly kosher.
\par
In writing so, we have lost longitudinal diffeomorphisms in both cases , but we have gained Weyl invariance. Weyl 
Ward's identity read in this case
\be
g^{\m\n}{\d S\over \d g^{\m\n}}\equiv 0
\ee
where
\be
\Theta_{\m\n}\equiv {\d S\over \d g^{\m\n}}=|g|^{1/n}\left({1\over n}\left(\nabla\phi\right)^2~g_{\m\n}-\nabla_\m\phi\nabla_\n\phi\right)
\ee
The point is that in spite of the fact that the action [\ref{TDS}] reduces in a free falling frame to the flat space one, the corresponding "energy-momentum tensor", $\Theta_{\m\n}$ fails to do so. Actually, only its traceless piece is recovered; the contribution of the potential has disappeared completely. This "energy-momentum" does not generate the conserved energy-momentum in flat space. What happens is that Rosenfeld's prescription \cite{Rosenfeld} needs the coupling to gravity to be full \text{Diff} invariant in order for it to work properly; that is to reduce to  Noether's current in flat space.
\par
Nevertheless, when gravity is dynamical according to [\ref{UG}] Bianchi identities ensure consistency (cf \cite{Redux} for a short review).
\par
The model can be easily modified to incorporate the Higgs mechanism, that is
\begin{widetext}
\bea\label{Higgs}
S_H&&=\int d^n x~\bigg\{|g|^{1/n}g^{\m\n}\left(\nabla_\m+i e |g|^{1\over 2n}~A_\m\right)\left(g^{1\over 2n}\phi^*\right)\left(\nabla_\n-ie |g|^{1\over 2n} A_\n\right)\left(|g|^{1\over 2n}\phi\right)-V(|g|^{1\over 2n}|\phi|)\bigg\}=\\
&&=\int d^n x~\bigg\{|g|^{2/n}g^{\m\n}\left[{1\over 2n}{\nabla_\m g\over g} \phi^*+\nabla_\m \phi^*+i e |g|^{1\over 2 n}~A_\m\phi^*\right]\left[{1\over 2n}{\nabla_\n g\over g} \phi+\nabla_\n \phi-i e |g|^{1\over 2 n}~A_\n\phi\right]-{\l\over 4!}~\left(|g|^{1\over n}|\phi|^2-v^2\right)^2\nonumber
\eea
\end{widetext}
where a quartic symmetry breaking potential has been incorporated in the last formula.

\section{Spontaneous symmetry breaking.}
Let us summarize. If we are willing to break the full \text{Diff} symmetry to its transverse subgroup, we can earn in compensation Weyl invariance. This gauge invariance does not however imply scale invariance in a free falling frame; otherwise Weyl invariance would not be  compatible with a nontrivial mass scale. That is in contrast with what happens in full Diff invariant theories \cite{Duff}\cite{Karananas}.
\par
In other words, when the symmetry is reduced to TDiff, Weyl invariance in curved space does not necessarily reduce to scale invariance in flat space.
\par
There are well-known notorious difficulties to break spontaneously any of these symmetries. Let us now study a (rather contrived) way to do that in our case.
\par
Consider two  symmetric tensor fields $f_{\m\n}$ and $\Lambda^{\a\b}$ and the following \text{Diff}-invariant term in a lagrangian
\be
L_T=\sqrt{|\text{det}~f_{\m\n}|}~\left[\Lambda^{\a\b}(x)\left(f^N_{\a\b} -g_{\a\b}\right)+ {\cal L}_{D}\right]
\ee
where ${\cal L}_D$ is a $\text{Diff}$-invariant lagrangian, and
\be
f^N_{\a\l}\equiv f_\a^\b f_\b^\g\ldots f^\s_\l\quad (\text{N  times})
\ee
\par
The EM for $\Lambda$ imply that
\be
f^N_{\a\b} =g_{\a\b}
\ee
and then
\be
f\equiv \text{det}~f_{\m\n}=g^{1\over N}
\ee
On $\Lambda$-shell then the symmetry has been reduced to the subgroup $H$
\be
L_T=|g|^{1\over 2 N}~{\cal L}_D
\ee
In that way the $\text{Diff}$ invariance will be broken to $\text{TDiff}$ invariance, which will be enhanced to Weyl symmetry if the power of $N$ is chosen properly.
Let us now work out a simple example of the breaking from  Diff to TDiff.
\bea
S=&&\int d^n x\bigg\{\sqrt{|\text{det}~f_{\m\n}|}~\left[\Lambda^{\a\b}(x)\left(f^n_{\a\b} -g_{\a\b}\right)+\right.\nonumber\\
&&\left.+{1\over 2}~g^{\m\n}\nabla_\m\phi\nabla_\n\phi\right]+{\l\over 4!}\left(\phi^2-v^2\right)\bigg\}
\eea
On shell this reduces to
\bea
S 1&&=\int d^n x\bigg\{|g|^{1/n} \left(-{1\over 2}~ g^{\m\n}\nabla_\m\phi\nabla_\n\phi\right)+\nonumber\\
&&+{\l\over 4!}~\left(\phi^2-v^2\right)^2\bigg\}
\eea
A spontaneous symmetry breaking solution can be found in the unimodular gauge where 
\be
|\hat{g}|=1
\ee
\bea
&&S=\int d^n x~\bigg\{ -{1\over 2}~ \hat{g}^{\m\n}\nabla_\m\hat{\phi}\nabla_\n\hat{\phi}+{\l\over 4!}~\left(\hat{\phi}^2-v^2\right)^2 \bigg\}
\eea
Then, through Weyl rescalings
\bea
&&g_{\m\n}\equiv \Omega^2 \hat{g}_{\m\n}\nonumber\\
\eea
we recover a whole Weyl orbit of solutions.
\par
A similar analysis can be easily be worked out for the Higgs model.
\par
\section{Conclusions.}
We have discussed Weyl invariant theories that are not scale invariant in a free falling frame, where the space-time metric reduces to the flat one. This is due to the fact that our theories are not invariant under all diffeomorphisms, but only under transverse ones.
\par
We have also discussed simple models in which the Diff invariance is spontaneously  broken to TDiff.
\par
A natural question at this point would be: What good is Weyl symmetry in this context and why would we want it?
\par
 The answer to that is manyfold. 
 \par
 For once, it forbids a cosmological constant, or rather, any constant term in the lagrangian does not couple to dynamical gravity. 
 \par
 Also the concept of a mass changes, insofar as it applies not really to the scalar field, but rather to the Weyl invariant combination $|g|^{1\over 2n}~\phi$. The scalar field $\phi$ has not physical meaning by itself.
 \par
 The final dream is of course that if Weyl invariance could survive at the quantum level, this theory would be a finite one \cite{Fradkin}, although
  the one loop calculations made up to now lead to  anomalies  in the Weyl Ward identities \cite{AlvarezGMHVM}. 
  \par
  More work is needed, however, in this and related issues.
\section{Acknowledgements}
We have enjoyed discussions with Mario Herrero-Valea and CP Martin.
 This work has been partially supported by the European Union FP7 ITN
INVISIBLES (Marie Curie Actions, PITN- GA-2011- 289442) and (HPRN-CT-200-00148);
COST action MP1405 (Quantum Structure of Spacetime), COST action MP1210 (The
String Theory Universe) as well as by FPA2012-31880 (MICINN, Spain)), and S2009ESP-1473 (CA Madrid).  This project has received funding from the European Union' s Horizon 2020 research and innovation programme under the Marie Sklodowska-Curie grant agreement No 690575. This project has also received funding from the European Union' s Horizon 2020 research and innovation programme under the Marie Sklodowska-Curie grant agreement No 674896.
The authors acknowledge the support
of the Spanish MINECO Centro de Excelencia Severo Ochoa Programme under grant
SEV-2012-0249.

\appendix

\newpage

\end{document}